# Rounded notch method of femoral endarterectomy offers mechanical advantages in finite element models

David Jiang[1], Dongxu Liu[1], Efi Efrati[2], Nhung Nguyen[1], Luka Pocivavsek[1]

[1]University of Chicago Medicine, Section of Vascular Surgery and Endovascular Therapy, Chicago, USA
[2]Weizmann Institute of Science, Department of Physics of Complex Systems, Rehovot, Israel

Corresponding Author:
David Jiang, MD
5841 S. Maryland Ave
O-234
Chicago, IL, USA 60637
David.jiang@uchicagomedicine.org

**Abstract**
Objective: Use of a vascular punch to produce circular heel and toe arteriotomies for femoral endarterectomy with patch angioplasty is a novel technique. This study investigated the plausibility of this approach and the mechanical advantages of the technique using finite element models.
Methods: The patient underwent a standard femoral endarterectomy. Prior to patch angioplasty, a 4.2 mm coronary vascular punch was used to created proximal and distal circular arteriotomies. The idealized artery was modeled as a 9 mm cylinder with a central slit. The vertices of the slit were modeled as: a sharp V consistent with traditional linear arteriotomy, circular punched hole, and beveled punched hole. The artery was pressurized to achieve displacement consistent with the size of a common femoral artery prior to patch angioplasty. Maximum von Mises stress, area-averaged stress, and stress concentration factors were evaluated for all three models.
Results: Maximum von Mises stress was 0.098 MPa with 5 mm of displacement and increased to 0.26 MPa with 10 mm of displacement. Maximum stress in the uniform circular model was 0.019 MPa and 0.018 with a beveled notch. Average stress was lowest in the circular punch model at 0.006 MP and highest in the linear V notch arteriotomy at 0.010 MPa. Stress concentration factor was significantly lower in both circular models compared with the V notch.
Conclusions: Femoral endarterectomy modified with the creation of circular arteriotomies is a safe and effective surgical technique. Finite element modeling revealed reduced maximum von Mises stress and average stress at the vertices of a circular or beveled punch arteriotomy compared with a linear, V shaped arteriotomy. Reduced vertex stress may promote lower risk of restenosis.

**Introduction**
Femoral endarterectomy has remained the gold standard surgical management technique for patients with atherosclerotic occlusive disease of the iliofemoral vessels and has demonstrated durable long-term outcomes[1–3]. Unfortunately, restenosis remains common following endarterectomy. Several techniques have been developed to reduce the risk of restenosis including the use of a bovine pericardial, expanded polytetrafluoroethylene (ePTFE) or other autologous patch[2,4]. The use of a patch is standard following femoral endarterectomy at most institutions; however, proper technique and sizing is critical to prevent an increased risk of stenosis due to patch angioplasty itself[5]. Prior studies have identified patches after carotid endarterectomy as areas of perturbed flow with resultant increases in oscillatory shear index and turbulence, which contribute to neointimal hyperplasia and restenosis[5–7]. Critically, however, there has been limited exploration of the role of the geometry of the arteriotomy created in the endarterectomy itself on subsequent restenosis.

A linear arteriotomy is created and extended proximally and distally in the traditional femoral endarterectomy and the resultant geometry has sharp "V" shaped notch at its vertices. The need for rounded arteriotomies is better described in the cardiothoracic literature where vascular punches have long been used for aortotomies prior to aortosaphenous vein end-to-side anastomoses during coronary artery bypasses[8–10]. The vascular punch is preferred in this situation because of its ability to create circular and symmetrical arteriotomies through the aortic wall with reduced need to trim or augment the arteriotomy. This technique was subsequently adopted by microsurgeons to create symmetrical end-to-side anastomosis arteriotomies especially in areas with atherosclerotic vessels. The vascular punch performed well in calcific vessels with minimal intimal disruption[8,9]. In applying the vascular punch to improve the current technique for femoral endarterectomy, the goal was to identify methods to reduce areas of stress concentration.

Stress concentration (*K*) is a well-known phenomenon in solid and fracture mechanics, which has also been applied into the vascular literature, predominately in understanding the impact of stent technology on the native vasculature[11,12]. Vascular stents produce non-native radial stress on the arterial wall to which the artery responds via adaptations to reduce this new stress concentration. Hypertension has been found to induce similar responses in arterial walls[13,14]. The response to stress concentration involves neointimal hyperplasia and thickening of the arterial wall in the area of stress concentration[14]. This is one proposed mechanism by which restenosis can occur following stenting. Creation of an arteriotomy for femoral endarterectomy produces a non-native geometry which similarly generates a non-native stress concentration at its vertices. Here, we describe a finite elements model (FEM) analysis of how the rounded notch method for femoral endarterectomy reduces stress concentration compared to traditional linear arteriotomy.

## Methods
### Operative Details
Informed consent was obtained from the patient for research purposes. Institutional review board approval was not required.

A 66 year old patient was initially diagnosed with peripheral arterial disease (PAD) several years prior to this operation. He subsequently underwent bilateral below-knee amputations and was able to ambulate using a prosthesis until six months prior to presentation when he developed a chronic, poorly-healing left lower extremity wound. Pre-operative duplex ultrasonography demonstrated peak systolic velocity (PSV) increase from 62 cm/s to 350 cm/s at the left common femoral artery and PSV increase of 72 cm/s to 218 cm/s at the left mid-superficial femoral artery consistent with hemodynamically significant stenosis. Diffuse calcific plaques were noted as well, which was consistent with computed tomography angiography (CTA) imaging. CTA demonstrated multifocal left lower extremity stenosis that was most severe at the common femoral and popliteal arteries. The left profunda femoris artery was densely calcified and the left superficial femoral artery was partially occluded (Figure 1).

The patient was consented and underwent a left femoral endarterectomy with patch profundoplasty. The operation began in the usual fashion with infrainguinal exposure of the left common femoral, superficial femoral, and profunda femoris arteries. Proximal and distal vascular control was obtained using a combination of silastic loops and vascular clamps. A linear arteriotomy was created on the anterior surface of the common femoral artery and extended onto the profunda femoris artery using Potts scissors. An endarterectomy was performed using the Freer elevator. Distal external iliac plaque was removed using an eversion endarterectomy technique. The vertices of the common femoral artery and the profunda femoris were then rounded using a 4.2 mm coronary punch. A bovine pericardial patch profundoplasty was then performed in the usual fashion using 6-0 polypropylene suture (Figure 2).

**Finite Element Modeling**
In this section, the deformation behavior of femoral endarterectomy models under pressure is simulated using Finite Elements Modeling (FEM). Three femoral endarterectomy models with different arteriotomy strategies were modeled. The deformation and stress distribution of these models were compared and analyzed.

*Geometry and Boundary Conditions*
The geometry, dimensions and boundary conditions of the computational models are shown in Figure 3. The artery was simplified as a cylinder with a 9 mm diameter. The length of the artery was 64 mm, and the linear, V notch, arteriotomy was created at the center on the anterior surface with a length of 32 mm. The thickness of the artery was assumed to be 0.6 mm. The dimensions of the computational models were approximated from patient data. The longitudinal movement of the nodes on the edges of both ends was constrained. To eliminate rigid body translation, the nodes at the bottom of the artery were also constrained in the tangential and radial directions (Figure 3). A pressure of 0.1 kPa was applied to the internal surface of the artery to simulate the loading of blood pressure. It was noted that the applied pressure was lower than normal blood pressure due to the lack of a patch in the arteriotomy.

Three arteriotomy strategies were adopted in this work: arteriotomies without punch (Figure 4a), with a circular punch (Figure 4b) and with a beveled punch (Figure 4c). In circular arteriotomy, the coronary punch creates uniform circular holes with a diameter of 0.8 mm at both ends of the arteriotomy. In the beveled model, two types of beveled holes were considered for comparison. The computational models were meshed by S4R elements. The mesh at the regions surrounding both ends of the arteriotomy was refined with a minimum element size of 0.01 mm to improve stress field resolution.

*Material Properties*
The femoral artery was modeled as a Neo-Hookean hyperelastic material. The mass density of the artery wall was assumed to be $1.12 \times 10^{-9}$ g/mm[15]. The initial shear modulus and the bulk modulus were assumed to be 0.02408 MPa and 2.4 MPa, respectively[16].

*Quantification of Opening Displacement*
The pressurization on the internal surface of the artery wall results in the deformation of the artery wall and the opening of the arteriotomy. An opening displacement was measured to quantify the degree of artery wall deformations and the effect of different arteriotomy strategies on the opening behavior. The opening displacement was defined as the distance between points A and B, which are located at the centers of the corresponding edges of the arteriotomy (Figure 5).

*Evaluation of Average Stresses*
The average stress of the tip region was calculated to analyze the effect of punching on stress distributions. The stress surrounding the vertex of the femoral arteriotomy was averaged over predefined zones. The distance from the vertex was used as the criterion to define the zones in the linear arteriotomy model, and the distance from the edge of the punch defect was used to determine the zones in the punched arteriotomy models (Figure 6). The distance ranged from

0.1 mm to 2.5 mm (Figure 6d). Nine zones were defined using this method for each model. The maximum distance from the vertex or edge was 2.5 mm to enclose the high-stress regions.

*Calculation of Stress Concentration Factor*
Stress concentration factor *K* is calculated to characterize and quantitate the degree of stress concentration in models with different arteriotomy approaches[17–19]. In the simplest system of an infinite plane under tension with an oval or circular hole in plane, $K_t = 1 + 2\sqrt{a/\rho}$, where $a$ is the width of the hole and $\rho$ the root radius, it becomes clear that small values of $\rho$ yield high $K_t$. Therefore, we expect that larger root radii reduce the stress concentration factor[19]. We applied a similar dimensionless ratio to this Neo-Hookean hyperelastic system as a means to quantify differences in stress concentration amongst different geometries. The stress concentration factor *K* was defined as the ratio of the maximum stress ($\sigma_{max}$) at the vertex or punched hole edge to the reference stress ($\sigma_{ref}$), i.e., $K = \sigma_{max}/\sigma_{ref}$. Therein, the reference stress $\sigma_{ref}$ is the average stress in the ring area which is marked in red in Figure 6d.

**Results**
**Clinical Outcome**
The patient was evaluated in the clinic one month post-operatively at which time he was recovering appropriately with general improvement of his symptoms. The previously chronic wound of the left lower extremity demonstrated signs of healing with new granulation tissue in the wound bed and no clinical signs of infection. Arterial duplex ultrasonography was performed; the left femoral endarterectomy site was patent. PSV at the common femoral artery was substantially lower at 68 cm/s compared with 350 cm/s pre-operatively.

**Finite Element Modeling Results**
The vessel models were pressurized until the original edges of the mesh achieved 5 and 10 mm displacement to mimic the transverse dimensions of a typical femoral patch angioplasty. Figure 7 plots the relationship between the opening displacement and the applied pressure. Under the same pressure, the opening displacements of the circular punch model and the beveled model overlap and are larger than the no-punch model. Figure 8 demonstrates the stress profile at the proximal vertex of the longitudinal femoral arteriotomy for a typical linear geometry and the modified punched-arteriotomy configuration with a rounded vertex modeled using circular or beveled end design. The maximum stress in the no-punch model concentrates at the vertex of the no-punch model, while the high-level stress surrounds the edge of the punched defects in the circular punch and beveled punch model. The evolutions of maximum von Mises stresses with opening displacement in different models are presented in Figure 9. Maximum von Mises stress was 0.098 MPa with 5 mm of displacement and increased to 0.26 MPa with 10 mm of displacement. In contrast, the maximum von Mises stress at the punched defect of the arteriotomies with a uniform circular and beveled punch with 10 mm opening displacement was 0.019 MPa and 0.018 MPa, respectively. The configuration of the punched model vertex did not significantly affect the stress profile at the tip.

Using the proposed method in section '*Evaluation of Average Stresses*', the average stresses in the models are calculated. Their evolutions with opening displacements are demonstrated in Figure 10-12. In all models, average stress was greatest in Zone I, 0.1 mm from the tip or edge of the punched defect, and lowest in Zone IX, 2.5 mm away. The average stresses in all zones of the no-punch model are higher than the counterparts in the punched models. At an opening displacement of 5 mm, the von Mises stress in Zone I of the linear arteriotomy was approximately 0.010 MPa (Figure 10) compared with 0.006 MPa in the circular punch model (Figure 11) and 0.007 MPa in the beveled punch model (Figure 12); similar trends were identified at 10 mm opening displacement. Averaged von Mises stress decreased with distance from the vertex in all models.

Figure 13 shows that the stress concentration factor in the no-punch model was obviously higher than that in other models. *K* increased monotonically with the opening of slit in the no-punch model, while it remained nearly constant in the punched models.

**Discussion**
Notches and apices are natural sources of non-homogenous stress concentration in a variety of systems spanning from theoretical infinite planes to the human vasculature[20]. In arteries, stress concentration in these regions is linked to homeostatic perturbations which initiate processes that cause arterial wall thickening and promote restenosis[11,14]. This response is an adaptive method by which arteries reduce areas of stress concentration[14]. Unfortunately, this innate response undermines the success of femoral endarterectomies by causing stenosis which can lead to non-lamellar flow and plaque reformation. The present study investigated the stress distribution and concentration in different geometries of arteriotomy models and revealed interesting findings regarding the effect of geometry on stress concentration and distribution.

The stress profiles at the proximal vertex of the arteriotomy models emphasizes the influence of geometry on stress distribution. In the no-punch model, the maximum stress concentrated at the vertex, while in the circular punch and beveled punch models, high-level stresses were measured across the arc of the punched defects. This demonstrates that the presence of a hole or punched defect redistributes stress over a broader area. Prior work has shown that the natural compensatory response of arteries seeks to achieve a similar goal[14]. The direct correlation of maximum von Mises stress with restenosis is not well-established. Nonetheless, von Mises stress has been shown in other systems to contribute to the development of restenosis[13,14,21]. The use of a vascular punch allows for a cleaner arteriotomy in practice and is hypothesized to minimize endothelial denudation and disruption compared to a sharp technique using Potts scissors which would yield smaller von Mises stresses. The smaller von Mises stresses in the rounded notch configurations may also contribute to reduction inflammation from mechanical stresses which can otherwise result in neointimal hyperplasia and stenosis as seen in stenting[11,13].

Analyzing the maximum von Mises stresses with increasing opening displacement revealed that the no-punch model exhibited higher stress values compared to the punched models. At 10 mm opening displacement, the maximum von Mises stress in the arteriotomy models with a uniform circular punch and beveled punch was significantly lower than that in the no-punch model. This implies that the presence of a larger root radius, as achieved through the punched defects, effectively reduces the stress concentration factor. The stress concentration factor calculation supported this observation, indicating that larger root radii lead to decreased stress concentration. This is consistent with Figure 13 wherein stress alleviation improved with larger opening displacement.

The evaluation of average stresses in different zones of the models further elucidated the impact of geometry on stress distribution. Across all models, the average stress was found to be highest in Zone I, near the tip or edge of the punched defect, and lowest in Zone IX, situated 2.5 mm away. This indicates that the stress levels decrease with increasing distance from the vertex in all geometries. Moreover, the average stresses in the punched models were consistently lower than those in the no-punch model, implying that the introduction of punched defects mitigates stress concentration.

Future work is needed to overcome several current limitations. A patient specific geometry would provide better resolution than the idealized arterial geometry used in the FEM. The idealized model does not capture key components such as branching of the profunda femoris artery relative to the common femoral artery nor does it account for the natural curvature of the femoral vessels. These attributes are likely less important when considering the stress analysis performed at the vertices; however, would become key components of future computational fluid dynamics (CFD) analyses. Addition of a patch object into the model and a more realistic simulation of patch insertion and unclamping would be necessary as well. It is important to note that the stress concentration factor approximation derived from FEM was applied to a Neo-Hookean hyperelastic material, not a linearly elastic material and was primarily used to understand relative differences in $K$ due to geometry changes

In conclusion, these findings highlight the importance of vertex geometry in patients undergoing femoral endarterectomy. The results demonstrate that larger root radii, achieved through punched defects, reduces stress concentration and redistributes stress to a larger area surrounding the defects which may reduce detrimental arterial remodeling and hyperplasia.

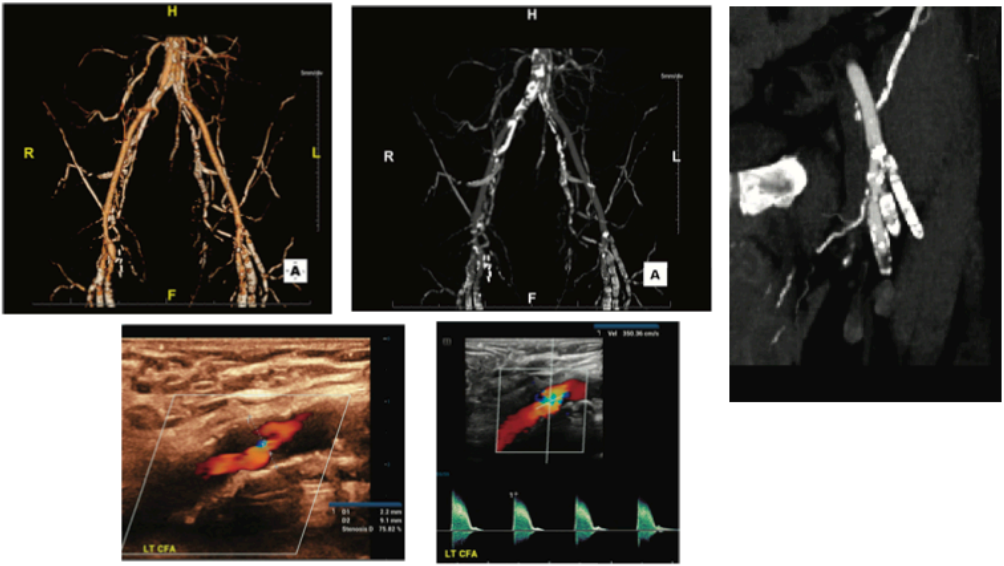

**Figure 1.** Pre-operative computed tomography (CT) and duplex ultrasonography (US) images demonstrating occlusive iliofemoral disease.

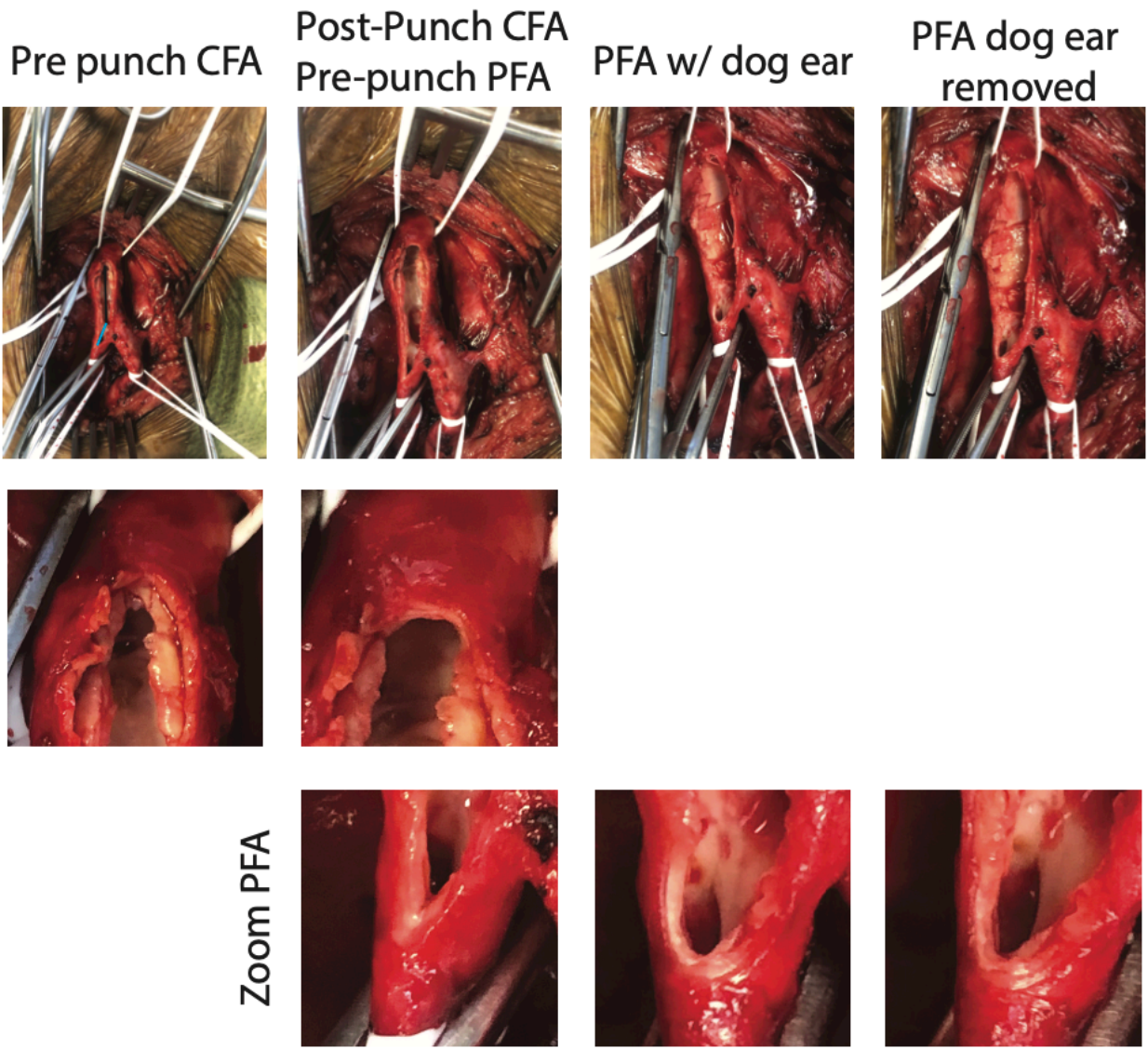

**Figure 2.** Intraoperative images demonstrate creation of the arteriotomy on the common femoral artery and its extension onto the profunda femoris artery. Zoom images highlight the rounded notch created using a 4.2 mm vascular punch.

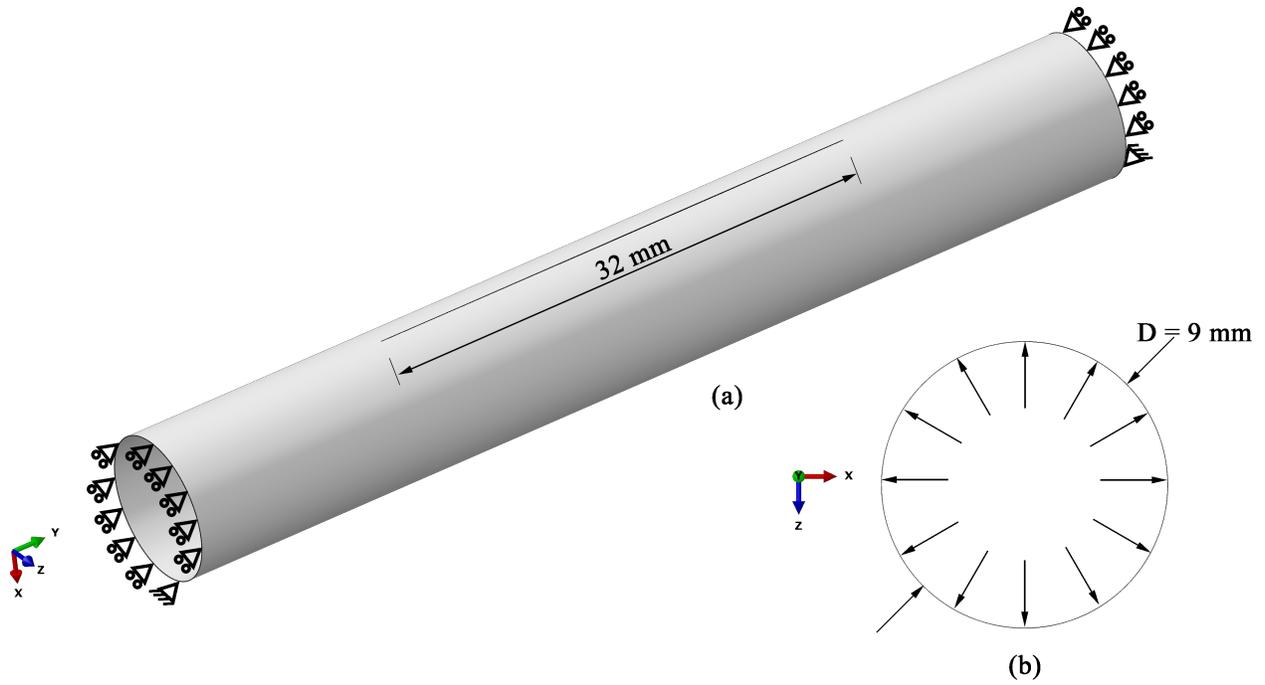

**Figure 3** (a) Geometry and boundary conditions, and (b) loading conditions (pressure) of the computational models.

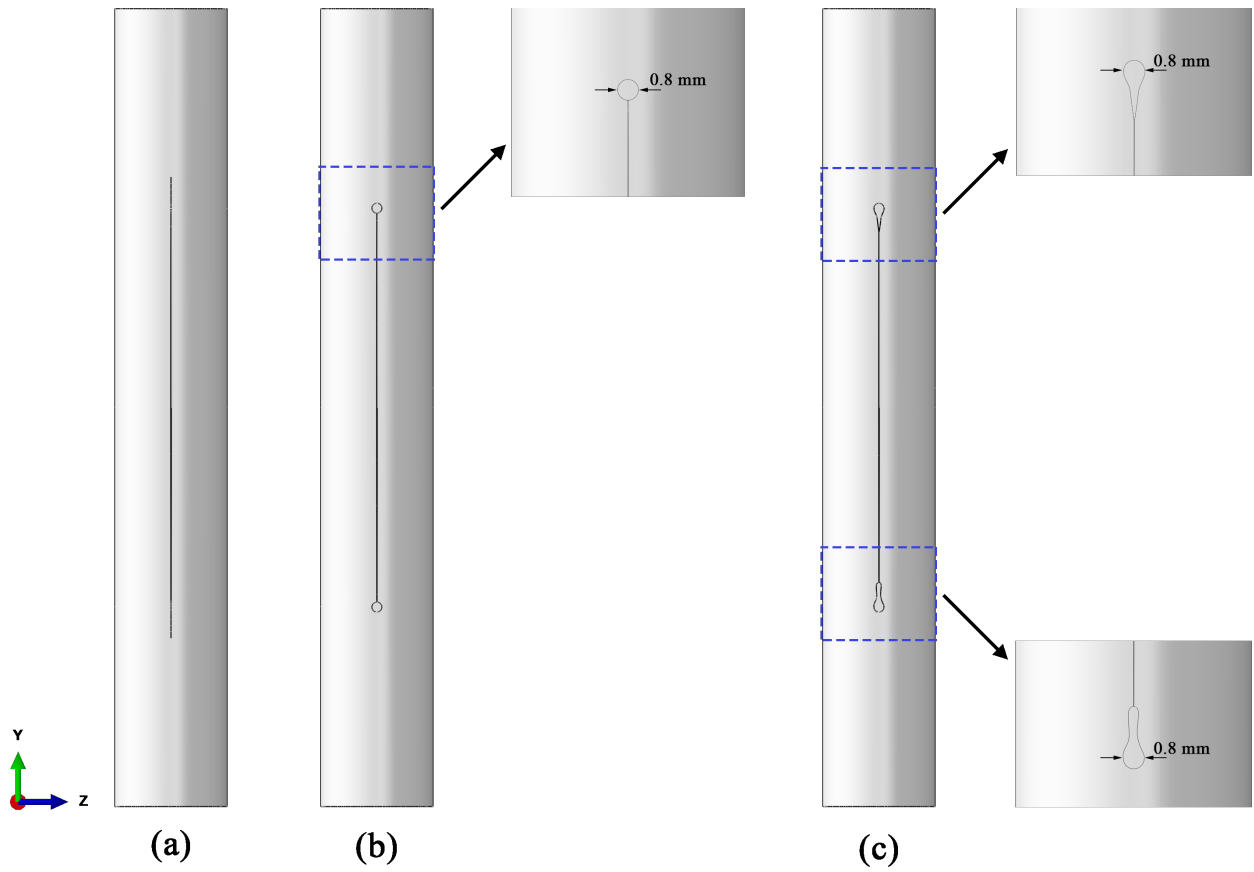

**Figure 4** Arteriotomy strategies: (a) no-punch model, (b) circular punch model and (c) beveled punch model.

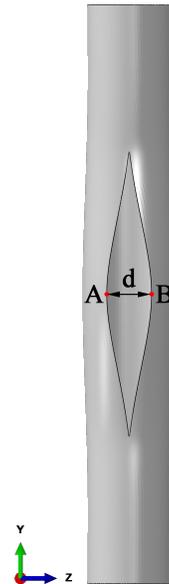

**Figure 5** Definition of opening displacement.

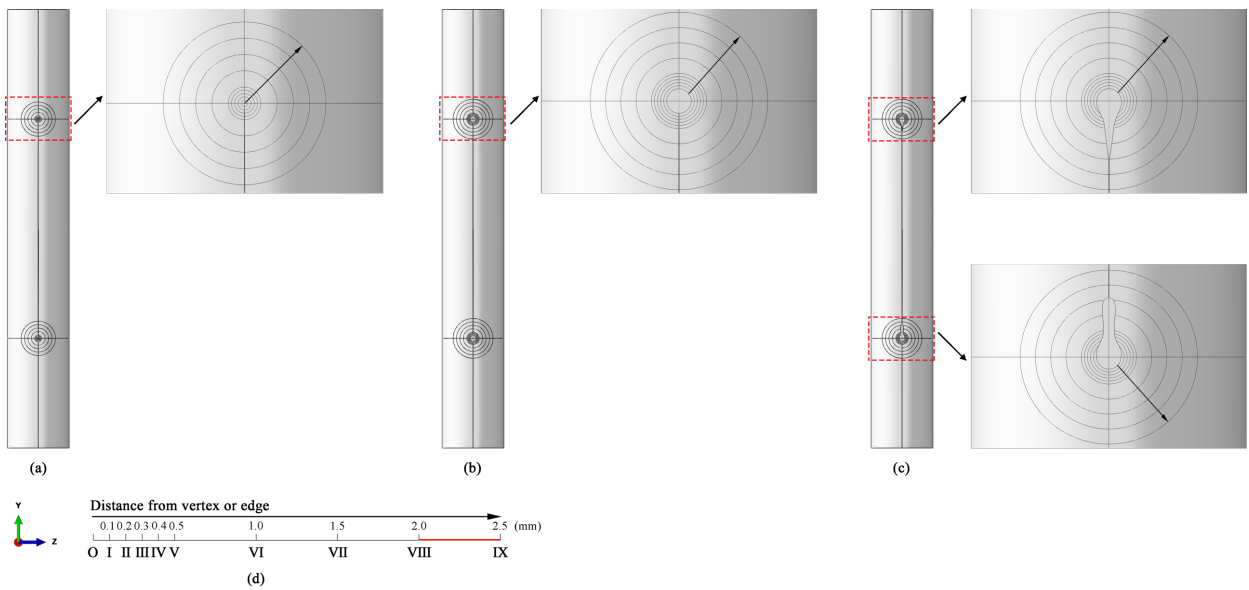

**Figure 6** Definition of zones for average stress calculation in (a) no-punch model, (b) circular punch model and (c) beveled punch model. Zone I to Zone IX in (d) correspond to the nine circular zones in (a), (b) and (c). The red zone marked in (d) between Zone VIII and Zone IX denotes a ring area for calculating reference stress $\sigma_{ref}$.

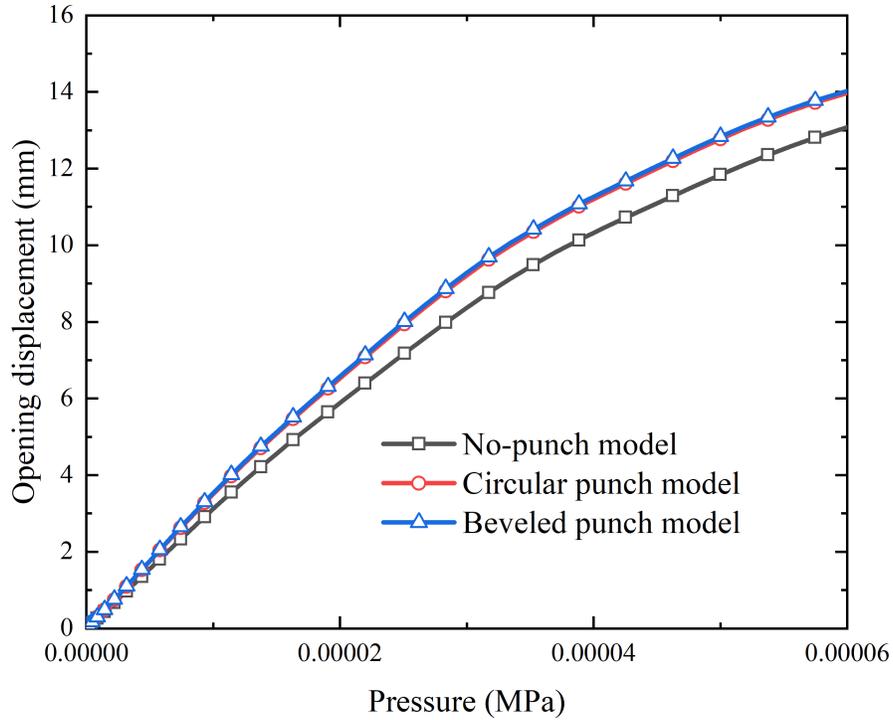

**Figure 7** Opening displacements in the no-punch model, circular punch model and beveled punch model.

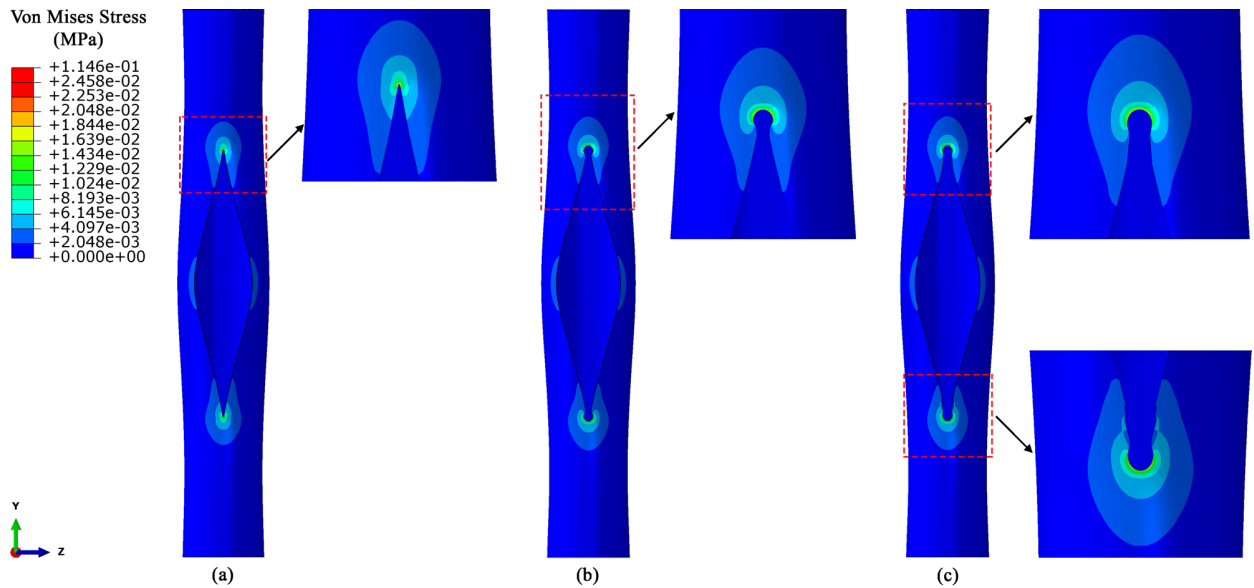

**Figure 8** Stress distributions in (a) no-punch model, (b) circular punch model and (c) beveled punch model under a pressure of 0.0235 kPa.

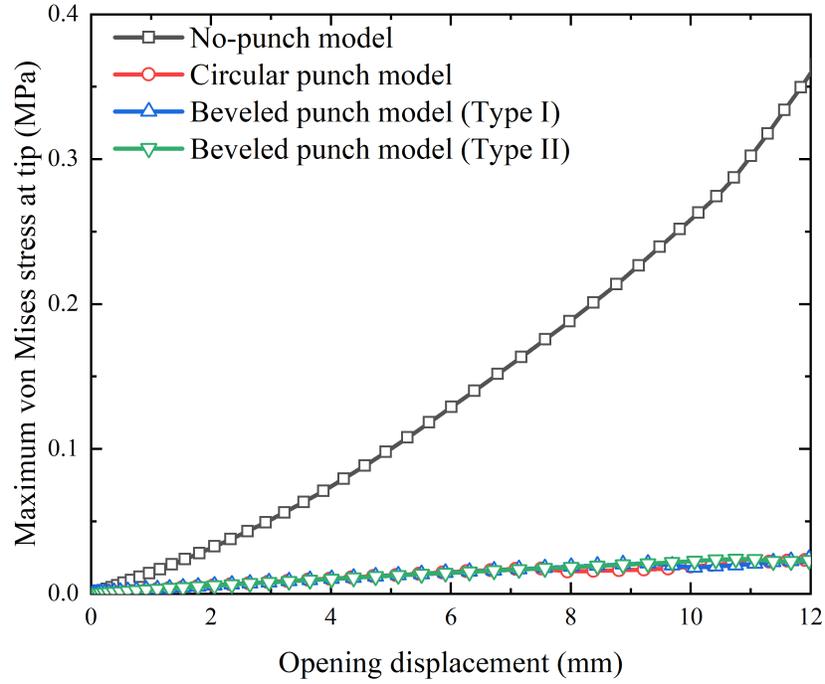

**Figure 9** Maximum von Mises stress in the models with different arteriotomy strategies.

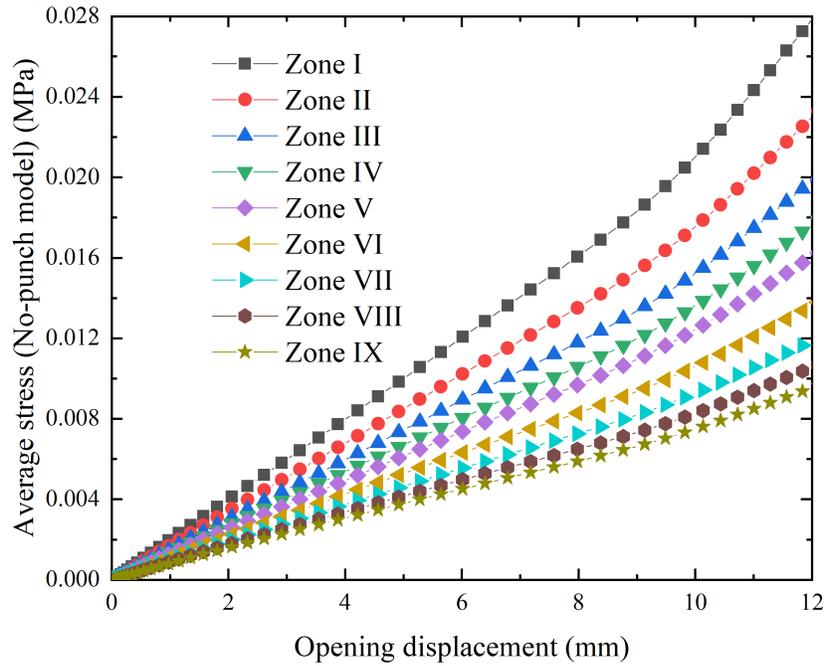

**Figure 10** Average stresses calculated over different zones in the no-punch model.

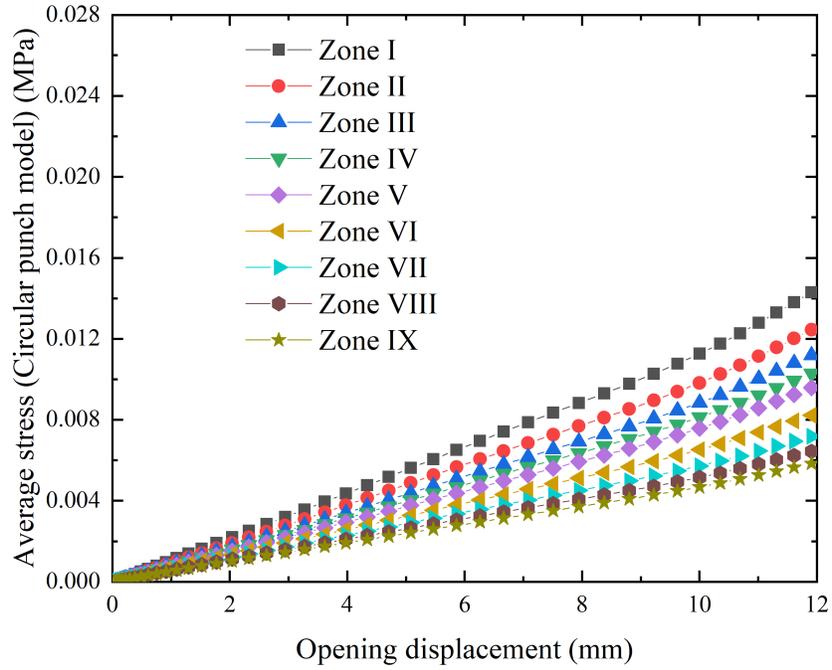

**Figure 11** Average stresses calculated over different zones in the circular punch model.

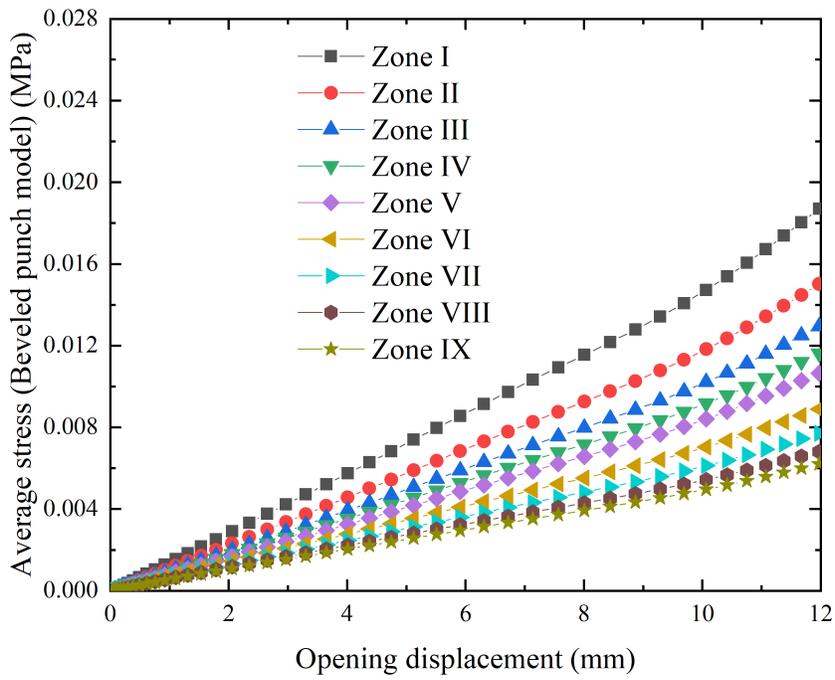

**Figure 12** Average stresses calculated over different zones in the beveled punch model (type I).

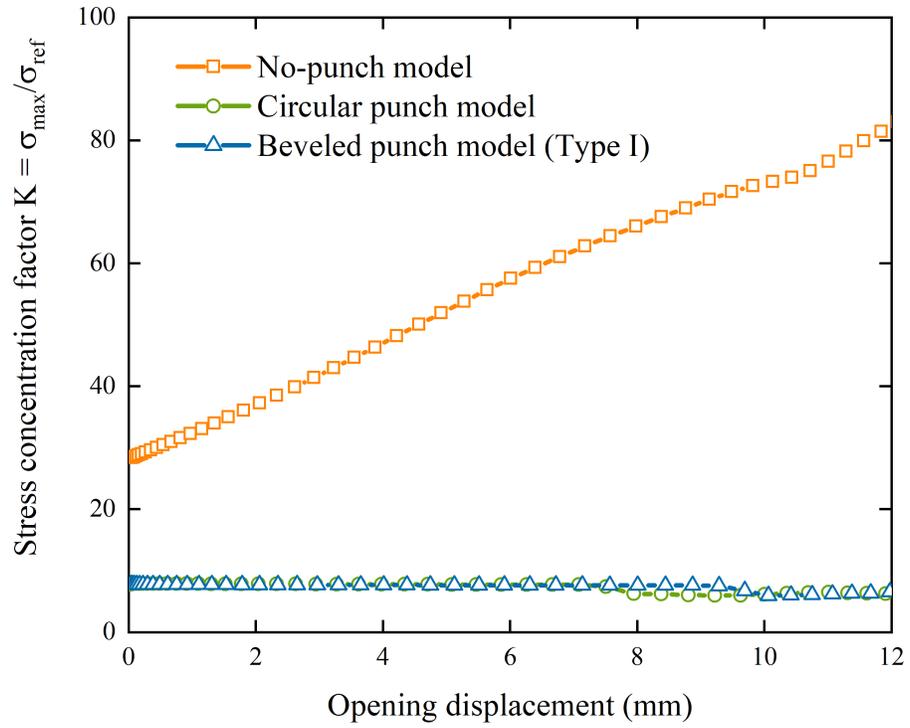

**Figure 13** Comparison of stress concentration factor in different models.